- *Type of the Paper (Article, Review, Communication, etc.)*

# Distributed Identity for Zero Trust and Segmented Access Control: A Novel Approach to Securing Network Infrastructure

**Sina Ahmadi**[1]

[1] The National Coalition of Independent Scholars (NCIS), sina0@acm.org

**Abstract:** "Distributed Identity" refers to the transition from centralized identity systems using Decentralized Identifiers (DID) and Verifiable Credentials (VC) for secure and privacy-preserving authentications. With distributed identity, control of identity data is returned to the user, making credential-based attacks impossible due to the lack of a single point of failure. This study assesses the security improvements achieved when distributed identity is employed with the ZTA principle, particularly concerning lateral movements within segmented networks. It also considers areas such as the implementation specifications of the framework, the advantages and disadvantages of the method to organizations, and the issues of compatibility and generalizability. Furthermore, the study highlights privacy and regulatory compliance, including the General Data Protection Regulation (GDPR) and California Consumer Data Privacy Act (CCPA), analyzing potential solutions to these problems. The study implies that adopting distributed identities can enhance overall security postures by an order of magnitude, providing contextual and least-privilege authorization and user privacy. The research recommends refining technical standards, expanding the use of distributed identity in practice, and discussing its applications for the contemporary digital security landscape.

## 1. Introduction

*1.1. Overview of Modern Cybersecurity Challenges*

Modern cybersecurity is rapidly developing as the threats become more varied and sophisticated [1]. Misleading links, stolen credentials, cryptocurrency payments, and robust internal network invasions have become part of the industry, challenging the capabilities of established security technologies. Credential theft is especially dangerous because it involves the attacker using a weak or hijacked password to gain access. Once inside a network, attackers can move laterally, gaining higher privileges necessary to reach valuable assets. Traditional security models, based on logical perimeters, struggle to address these threats, since they all involve implicit trust. This approach exposes much control to users or devices within the network, making organizations vulnerable to internal attacks and advancing persistent threats. Consequently, there is an urgent need for innovative security solutions that address these vulnerabilities and provide a robust defense against modern adversaries.

*1.2. The Rise of Zero Trust Architectures*

New security models have emerged in response to the shortcomings of traditional security models, and among them stands the Zero Trust Architecture (ZTA). The ZTA model does not trust any form of default trust, and its core model is encapsulated by the principle of "never trust, always verify" [2]. This model assumes that all the users, devices, and applications present are insecure and will need continuous authentication and authorization. ZTA is built on three fundamental principles: verify explicitly, apply least privilege access, and assume breach. In verification, the access request is authenticated using multiple factors about the user and the device, such as the user's identity and the device's health. The principle of least privilege grants users access only to the resources they need to perform their tasks, minimizing vulnerabilities. However, the assumption of breach pushes organizations to establish structures that reduce leakage in the wake of infiltration. The core of the ZTA approach involves logically separating network components and resources to limit the attack surface, allowing only authorized identities to access specific areas in real time. This architecture enhances security by integrating real-time monitoring and analytics to proactively detect and mitigate potential threats. ZTA also emphasizes encryption and segmentation to safeguard sensitive data, ensuring robust protection against internal and external cyber threats.

*1.3. Introduction to Distributed Identity*

Distributed identity represents a major shift in digital identity management, providing a decentralized model that aligns with ZTA principles. Unlike conventional systems, which rely on centralized data centers, distributed identity allows individuals to own and manage their identity information [3]. This approach leverages Decentralized Identifiers (DIDs) and Verifiable Credentials (VCs) to achieve secure authentication without compromising the individual's privacy. A user ID can be verified using cryptography without endangering the user's information. Therefore, some advantages of distributed identity include enhanced security due to the low utilization of central databases, beneficiaries' autonomy control, and enhanced stability due to the lack of critical control points. When augmented in ZTA, distributed identity strengthens network through least privilege access, continuous access validation, and minimizing reliance on privileged credentials. This integration allows organizations with private user identities, stronger security, and a robust defense shield against emerging threats, making distributed identity an essential component of contemporary cybersecurity.

*1.4. Research Scope and Objectives*

This research examines the incorporation of distributed identity into ZTA frameworks to address key challenges in modern cybersecurity. The main research question is how distributed identity can enhance network segmentation, prevent lateral movement, and manage exposure to credential-based threats. This research will analyze the technical and operational viability of implementing distributed identity along with ZTA, focusing on the use cases and advantages that both DIDs and VCs offer for fine-grained access control.

The main research questions address the enhancement of ZTA and the related issues of integrating distributed identity features, the key technological layer for their support, and possible challenges organizations may face, including interoperability, scalability, and user adoption. Furthermore, the research will investigate possible approaches to integrating distributed identity into existing networks to achieve perfect synchrony with the ZTA reference model with improved security and privacy. This informative research provides practical implications for enhancing the disidentified distributed identity as a revolutionary intervention in today's cybersecurity environment. The result will enable organizations to create robust, safe, and privacy-conscious networks, enhancing their ability to contain and avoid a breach and the loss of trust, including sensitive information in the evolving cyber threats environment.

## 2. Literature Review and Background

*2.1. Evolution of Identity Management*

The identity management systems are not what they used to be decades ago due to the growing concern of authorizing users and devices in a diverse digital environment [4]. Traditionally, identity management used a reference point or a particular database that all users would refer to, like the corporate Active Directory or a central IDP. These centralized systems have been widely utilized as the infrastructure for identity management across enterprises, enabling user access to resources according to the roles and credentials provided. However, as organizations and their networks evolved, managing identities centrally started having its own set of issues, including scaling, data leakage, and a dependency on a single point of failure. Centralized models also presented privacy concerns, as they stored vast amounts of sensitive personal data in a single location, making them attractive targets for cybercriminals.

Due to various problems associated with central joined identity systems, distributed joined identity systems were developed, which allowed many organizations to keep information about one unique user across different domains. This is done using Single Sign-On (SSO) and Security Assertion Markup Language (SAML), which makes it easier to move through the systems [5]. It enhances the user experience by preventing users from logging in multiple times to different services and increasing security through the trust established between identity and service providers. These trust relationships make sure that only authorized users will be allowed to gain access to these sites. However, as with the federated identity, it has its advantages of being convenient, secure, uncomplicated, and impracticalities involving the IDPs, which are central points of control but prone to being compromised by hackers.

The latest advancement in identity management is the distributed identity, which uses decentralized technologies to enable secure and private identity management. Distributed identity leverages distributed identifiers (DID) and verifiable credentials (VC), by which an individual owns his/her identity data and is not dependent on centralized authorities [6]. Distributed identity systems leverage any blockchain or distributed ledger to store identity data. This allows the user to completely control his/her digital profile and prevent identity theft, fraud, or privacy violation. Technologies that provide security features that align with this paradigm include blockchain, given its immutability, transparency, and tamper resistance, which can prevent unauthorized access or alteration of personal data.

| Aspect | Centralized Identity | Federated Identity | Distributed Identity |
|---|---|---|---|
| Control | Central authority | Shared among entities | User-controlled |
| Scalability | Limited by central infrastructure | Moderate | High |
| Privacy | Vulnerable to breaches | Improved but still central-dependent | Strong, minimizes data sharing |
| Resilience | Single point of failure | Multiple trusted entities | No single point of failure |
| Example Technologies | Active Directory, LDAP | SSO, SAML | DIDs, VCs, Blockchain |

**Table 1**. Comparison of Centralized, Federated, and Distributed Identity Systems.

*2.2. ZTA and Segmentation*

ZTA is a cybersecurity framework that operates on the principle of "never trust, always verify [7]."Unlike more traditional models that assume the user or device, once inside the perimeter, is trustworthy, ZTA expects the user or device may be malicious, whether internal or external to the network. This approach conflicts with traditional conventional thinking, whereby access control is attained through firewalls and other perimeter security. However, in ZTA, users and their devices are constantly validated at every step to grant access to sensitive data. Equation 1 demonstrates how segmentation quantifies risk reduction. In this equation, $R_{reduced}$ represents the reduced risk level,

$R_{baseline}$ denotes the baseline risk in traditional security models and $S$, indicates the segmentation factor.

$$R_{reduced} = R_{baseline} \times (1 - S) \qquad (1)$$

The core principles of ZTA include explicit verification, least privilege access, and assumed breach. This means there must always be some type of authentication and authorization of access request irrespective of the request's origin for any resource. This encompasses using Multi-Factor Authentication (MFA) and verifying the device's security status. Least privilege access allows the minimum access required to complete a task by a user and a device, thus offering minimal exposure to hostile insiders [8]. Lastly, unlike traditional security models that assume that external threats are kept at bay and will never get inside the network, ZTA supposes the opposite and implements controls that confine whatever got in, including its ability to move around laterally.

Network segmentation plays a critical role in zero-trust architectures. The use of subdomains in a network separates the network into different compartments, which, if an attacker infiltrates, they will not have easy access to other compartments [9]. This kind of segmentation is one of the low-level mitigations that minimize the attack surface and combat lateral movement, which attackers widely utilize to elevate their privileges and gain access to other systems. Segmentation only affords certain classes of assets, and if one segment is compromised, the breach does not spread all over the network.

| Component | Purpose | Example |
| --- | --- | --- |
| Verification | Authenticating access requests | Multi-Factor Authentication (MFA) |
| Least Privilege | Minimizing access rights | Role-Based Access Control (RBAC) |
| Assume Breach | Containment strategies | Network Segmentation, Micro-segmentation |
| Continuous Monitoring | Detecting anomalous behavior | SIEM, Behavior Analytics |

**Table 2.** Network Segmentation.

### 2.3. Distributed Identity in Practice

Some distributed identity systems started receiving attention in different fields, especially sectors that highly value privacy and security. Hyperledger Indy is one of the technologies that help implement distributed identity, a distributed ledger for building decentralized identifier systems [10]. Indy is a Hyperledger project that supports distributed infrastructure for identity. It applies the concept of blockchain to enable individuals to own global, safe, and authentic online identities. Companies adopting Hyperledger Indy can support the decentralized relations of users and services without the intermediation of other parties and give users complete control over their identity and information.

Another platform in the distributed identity area is Sovrin, which is based on the Hyperledger Indy. Sovrin is a clean slate decentralized network built for the creation, presentation, revocation, and validation of verifiable credentials (VC), thus making it easier for organizations to transition to distributed identity securely and in a scalable manner [11]. Sovrin also decentralizes its architecture which will reduce data silos and possible risks of identity fraud because it stores data centrally. Thus, Sovrin employs blockchain technology to provide seamless decentralization of identity credentials that cannot be altered, forged, or duplicated without permission or authorization. This devolved model simplifies the identity verification process, making it easy for organizations to extend secure and efficient access to resources. Sovrin has the potential to offer a self-sovereign identity model that allows individuals to reclaim control over credentials and increase privacy measures and overall risks of centralized identity systems. For this reason, Sovrin becomes insistent in the progressing paradigm of distributed identity.

In practice, distributed identity is used successfully in numerous applications within enterprises and sectors of critical infrastructures. For instance, in the financial services area, banking and other institutions are looking into using distributed identity systems to enhance efficiency in adoption and identity checks and balances amid related perils such as ID theft [12]. In decentralized identifiers, customers can prove their identity and transact with cryptographic provenance without compromising personal data. Similarly, in healthcare, distributed identity can enhance patient records' privacy and security, noting that patients would own and selectively share their health information only with healthcare providers/organizations as required in line with emerging healthcare privacy and data protection laws such as HIPAA and GDPR.

Distributed identity is also expected to enhance IoT security by providing a more secure way of authenticating devices within a highly connected network. Due to the absence of proper IT solutions for such devices, the IoT ecosystem rigs are usually exposed to attacks. Distributed identity creates a way of allowing only genuine devices to have entry to specific data, which makes IoT networks more secure [13].

| Feature | Hyperledger Indy | Sovrin |
| --- | --- | --- |
| Focus | Decentralized identity framework | Self-sovereign identity network |
| Underlying Tech | Blockchain | Blockchain |
| Scalability | Limited by current tech | High with the adoption of off-chain methods |
| Adoption | Open-source community-driven | Proprietary and community-driven |
| Key Strength | Customizable and flexible | Standards-aligned, easy integration |

**Table 3.** Comparison of Hyperledger Indy and Sovrin.

*2.4. Gaps in Current Research*

Despite the ability of distributed identity and ZTA frameworks being widely understood today, there are still areas with limited understanding. Another key issue is the absence of effective Solutions for Distributed Identity Combined with ZTA Concepts. While distributed identity and ZTA offer a solution to different facets of security, their joint advantages have not been fully optimized. There are few studies concerning how distributed identities might fit into existing ZTA frameworks and what might be the best integration approaches applicable in a large-scale enterprise context where old structures and frameworks create integration issues.

Another gap in the literature is the lack of solutions for the large-scale deployment of distributed identity. On the one hand, the advantages of decentralized identity management are quite evident; on the other, the obstacles that may become critical when considering put in practice remain unmeasurable. Barriers like lack of compatibility between distributed identity systems, legacy IT systems and structures, and overall awareness about decentralized ID management are significant challenges that must be overcome. However, there are certain concerns with scaling distributed identity systems with large organizations or governmental bodies where the amount of data and users is significantly large.

Furthermore, privacy issues have been raised again, mainly regarding how much information is safe or can be anonymously released to the public. Thus, distributed identity offers more control to the user. However, the issue of achieving the right balance between private, secure, and usable remains a challenging task that is still under investigation. It is also necessary to have more formalized processes to increase compatibility between spheres of application and create favorable conditions for the adaptation and implementation of these technologies.

## 3. Problem Definition

The lack of trust and access control are crucial issues in traditional security systems because most assume that trust is implicit at the center of their systems [14]. In these systems, users are usually given broad privileges based on the user's identity or role, which is dangerous when a hacker gets hold of these credentials or uses poor forms of authentication. Furthermore, the management of credentials in traditional systems is inconvenient and vulnerable to attacks, which suggests that there may be no control over the information exchanged. In these contexts, trust arrives after the user logs in and thus leaves systems vulnerable to horizontal movement and unauthorized access.

Integrating distributed identity with zero-trust architectures presents several barriers, both technical and organizational. From a technical perspective, the main obstacles are cross-platform integration of the distributed identity platform with legacy systems and its ability to accommodate many users and transactions. DIDs and VCs are used in distributed identity management, and they have to be incorporated into various systems that a modern organization employs, which can only be done by redesigning existing processes and IT security measures [15]. Moreover, challenges in integration between multiple identity management solutions and integration with old systems can greatly hinder the implementation process.

Organizational barriers are another factor that keeps pushing the organization backward in implementing new identity management perspectives. These challenges relate to the user adoption of distributed identity systems, where users and employees must be trained to use distributed identity systems and resist changing from a centralized identity model. It is also important for organizations to ensure that their employees take some training to avoid the great insecurity that comes with using these systems [16]. Due to these challenges, there is a compelling argument for a new approach that embraces the tenets of distributed identity in conjunction with ZTA.

## 4. Research Agenda

The purpose of the presented study is to assess the possibility of adopting distributed identity within ZTA frameworks to improve security, privacy, and authorization mechanisms in contemporary networks. The first goal is to examine the feasibility of this integration at a technical and operational level, focusing on factors like integration, complexity, and security. Further, the research aims to identify measures to address the adoption challenges, such as user training, organization-wide adoption, and integration of new technology infrastructures [17]. The study will offer practical recommendations to ensure distributed identities can be brought to mainstream adoption across ZTA by overcoming these challenges.

This study will use a research approach that includes a thorough review of existing literature, case studies, and technical frameworks to accomplish these objectives. This review will decompose top practices, conclusions, and misunderstandings related to the usage of distributed identity systems. The research will also compare the current identity management solutions and evaluate the changes when implemented under ZTA systems. The outcomes will define the efficiency of these technologies in protecting the network infrastructures, implementing access control, and strengthening cybersecurity. Therefore, through examining actual use cases and technical designs, the research will uncover how distributed identity can revolutionize cybersecurity practices.

## 5. Discussion

### 5.1. Security Benefits of Distributed Identity

Distributed identity is a significant paradigm shift for organizations to handle identity and access management data [3]. Another advantage of distributed identity is that it strengthens the forms of authentication using Decentralized Identifiers (DIDs) and Verifiable Credentials (VCs). The current identity systems require an intermediary, meaning an attacker can try to penetrate this authority. On the other hand, distributed identity democratizes this process, and users can manage their identity. This shift improves authentication by providing cryptographic proof of identity, which can be validated without decentralized storage or management. When sharing personal information with apps, the user can share only those parts of their identity, which can be dangerous, reducing the amount of information that can be exposed and the size of the attack [18].

Moreover, associating distributed identity with ZTA can minimize the attacker's movement within the network. Conventionally, these systems allow anyone access to almost all resources once a user's credentials are validated, and this allows attackers to ferry within the organization once they get hold of a username and password. However, with distributed identity, the authentication mechanism is linked with the particular access request, and it will determine permission by the roles and behavior in the context of real-time [19].

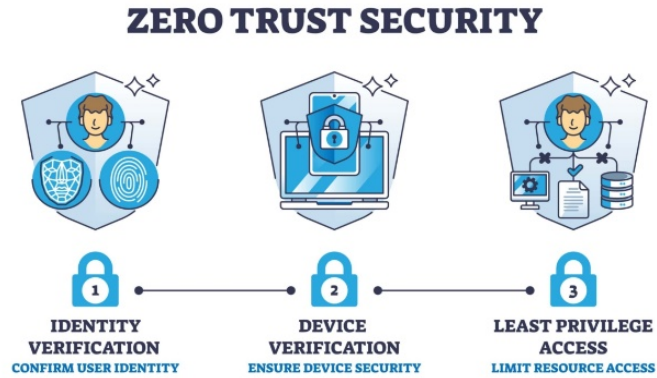

**Figure 1.** Distributed Identity with ZTA.

This results in decreased lateral movement and, in turn, an enhancement of the network segmentation since access requests can be constantly validated and authorized. In ZTA, any access request is considered to be coming from an untrusted entity, even if the user is inside the enterprise network [20]. When users are authenticated each time access is granted based on their identity and contextual factors, distributed identity enhances ZTA's least privilege access model to mitigate insider threats and outside attacks more effectively. Equation 2 describes distributed identity authentication mechanism with respect to access evaluation. $E_{Access}$ represents the access validation score which is calculated by average of the probability of successful authentication times the probability of meeting privilege requirements .

$$E_{Access} = \frac{\sum_{i=1}^{n} P_{auth}^{i} \times P_{privilege}^{i}}{n} \qquad (2)$$

*5.2. Practical Considerations*

The main advantages of integrating distributed identity into the ZTA model are obvious regarding security. However, organizations have gone through several practical questions to define how to use this solution. A major technical precondition for deploying distributed identity is the compatibility of decentralized identity solutions with existing systems [21]. Distributed identity empowers blockchain and distributed ledger technologies, including decentralized identifiers (DID) and verifiable credentials (VC). To achieve these technologies, organizations must confirm whether their current strategies are compatible with these technologies or whether organizations will need to adapt new platforms that favor interoperability between the current centralized models and the emerging decentralized models. Platforms that enable this integration include Hyperledger Indy, Sovrin, and other decentralized identity solutions.

Another technical characteristic that could be considered critical is the issue of scalability to accommodate several organizations. Distributed identity systems can be expected to accommodate many users and authentication calls and must not be slow to process them [22]. Despite being regarded as a highly secure technology, blockchain-based solutions may have issues connected to throughput and speed, especially when many transactions are involved. To manage this, there is the need to incorporate truly scalable consensus mechanisms and off-chain ledgers that will achieve better security and performance in the network.

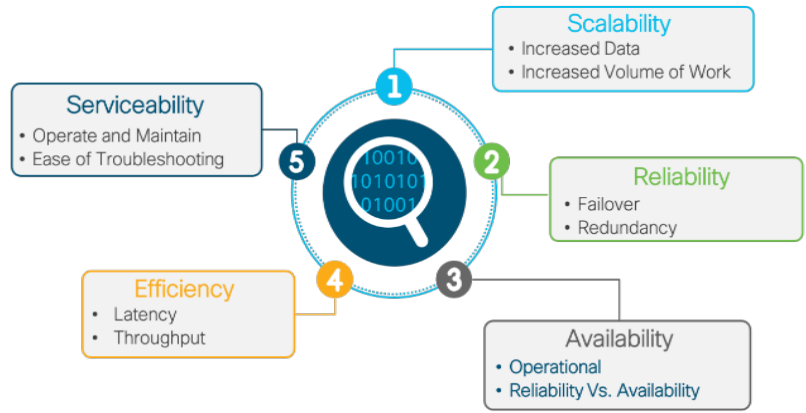

**Figure 2.** Scalability in Distributed Identity System.

From an economic perspective, there is also a cost-benefit analysis that organizations have to make before opting for distributed identity [23]. The long-term gains of improved security, decreased fraud, and users' power over their identity data outweigh the challenges. However, the costs of migrating to a distributed identity system are high. Such costs may include developing new infrastructure, training its employees, and system integration. However, the benefits of cutting initial costs are balanced by the potential for long-term savings, such as decreased rates of data breaches, better adherence to privacy legislation, and decreased administrative costs.

| Cost Component | Distributed Identity | Centralized Identity | Federated Identity |
|---|---|---|---|
| Initial Setup | High | Low | Moderate |
| Maintenance Costs | Moderate | High | Moderate |
| Risk Mitigation Costs | Low | High | Moderate |
| Compliance Costs | Low | High | Moderate |
| Overall ROI | High (long-term) | Low | Moderate |

**Table 4.** Comparison Between Distributed, Centralized and Federated Identity.

*5.3. Addressing Challenges*

However, there are several concerns that organizations need to deal with in their efforts to adopt distributed identity in cybersecurity. The greatest challenge of integrating decentralized identity with other systems is interoperability issues. It is crucial that distributed identity platforms and technologies being developed, such as blockchains, Distributed Identity Documents (DID), and Verifiable Credentials (VCs), have to interoperate with each other and legacy systems. To achieve this, standardization is vital. Standardization is an important prerequisite in ensuring that distributed identity systems can operate across platforms and ecosystems, including the W3C Verifiable Credentials and Decentralized Identifiers [24]. Organizations may also require essentially incorporating middleware or integration layers to connect organizations' decentralized identity solutions to other conventional systems. Equation 3 calculates the interoperability factor indicating the system's ability to function across heterogenous platforms. $C_j$ and $S_j$ reflect compatibility and scalability score of component j with distributed identity frameworks, in a system with $m$ components.

$$I_{interop} = \frac{\sum_{j=1}^{m} C_j \times S_j}{m} \qquad (3)$$

The issue of scalability also persists as an issue of great concern, especially given the large organizational structures that may have thousands or even millions of users. Distributed identity solutions, especially those based on blockchain, may encounter problems with the throughput and latency of transactions that could slow down decision-making related to access control. Some of these scalabilities can be solved by layer 2 scaling, where transactions are moved to a side chain, but the main chain remains secure and permanent. In addition, organizations can implement distributed identity integrated with existing centralized structures to benefit from both models.

In addition to technical challenges, user education and engagement strategies are critical for successful adoption [25]. As distributed identity changes traditional methods of identity management and control in users, organizations must ensure they offer proper training on the new systems. Introducing users to distributed identity and the associated advantages like privacy and sovereignty over personal information is crucial.

*5.4. Ethical and Legal Considerations*

With organizations embracing distributed identity solutions, discussing the legal and moral issues of decentralizing identity is critical. Regarding the implications of distributed identity, the most crucial issue is privacy. While decentralization of identity data empowers users to own their data and be in control of it, it raises key questions regarding the use, storage, and sharing of such data. Privacy preservation is the other critical principle, especially in distributed identity systems where data minimization and user consent guarantee privacy [26]. Users should be able to decide which credentials they want to reveal to others at a certain time when the demand is probably needed.

Also, distributed identity systems must adhere to some current data protection laws, including the General Data Protection Regulation (GDPR) in the European Union and the California Consumer Privacy Act (CCPA) in the United States. Some of these regulations highlight the user's rights, such as the right to access, the right to rectification, and the right to erasure. Equation 4 quantifies the level of privacy preservation in distributed identity system, where $D_{shared}$ is the amount of data shared during identity verification or an access control process and $D_{total}$ is the total data available about the user in the system.

$$P_{privacy} = 1 - \frac{D_{shared}}{D_{total}} \qquad (4)$$

## 6. Future Trends

*6.1. Emerging Technologies*

The application of distributed identity in the future heavily depends on the technologies in their development stage, especially AI and ML [27]. These technologies can go a long way in improving the flexibility and functionality of distributed identity systems. The technologies of AI and ML can study access patterns, users' behavior, and authentication requests and apply the results for adaptive security correction in real time. When harmonized with AI, distributed identity systems could respond to variations in security and identity verifications, depending on location, device, or the time of access attempted.

Moreover, future enhancements of blockchain-based and decentralized approaches promise to enhance such aspects of distributed identity management as scalability, security, and efficiency. Blockchain generally creates a reliable database to guarantee the credibility of digital identities and verifiable credentials (VCs).

*6.2. Policy and Standardization*

With distributed identity evolving into a global trend, standardization will be instrumental in promoting its uptake and compatibility across the globe. One of the key standards already being developed is W3C Decentralized Identifiers (DID), which will create a conceptual framework or more centralized identifiers [28]. This standard helps make it compatible with different distributed identity systems, meaning users can easily use their digital identity in another service or platform.

Similarly, there is expected to be a need for policy shifts if distributed identity is to grow beyond the pilot. Authorities and agencies worldwide have started to acknowledge the possibilities of decentralized identity for increased protection and privacy and the reduction of threats such as data breaches and identity thefts.

*6.3. Expanding Use Cases*

Distributed identity has the potential to revolutionize virtually every industry where there are currently issues with security, privacy, and efficiency. In the technology field, distributed identity systems would help simplify patients' access to their records while concurrently maintaining the security of the records, only permitting authorized individuals to review or transfer health information. Using credible credentials, healthcare-related companies and medical staff can guarantee that physicians and patients can surface pertinent records without violating the agreed privacy and HIPAA laws [29].

Distributed identity can help minimize fraud in the finance sector since users can provide more compelling identification information for transactions and account log-ins. It can also increase the efficiency of Know Your Customer (KYC) procedures, enabling financial organizations to safely validate their customers without having to depend on centralized databases or hackable legacy frameworks. The Internet of Things (IoT) is another area that can benefit significantly from distributed identity, given the nature of things connected to the internet.

## 7. Conclusion

This research underlines the value of distributed identity in promoting zero-trust architecture and compartmentalized access control. Distributed identity is a model of identity solution enabling users to take control of their identity data with less or no dependence on large databases. DID and VCs also support distributed identity that improves authentication solutions, privacy, and security threads, including lateral movements and credential-based threats. Incorporating this model into ZTA paradigms enables corporations to enforce adaptive and granular access controls to achieve safe and private interactions inherent in their network architectures. This is in line with the ZTA model of assuming a breach and requiring the user to prove the legitimacy of each interaction, enhancing the security of digital systems.

Specifically, the following suggestions can be made based on the exploration of distributed identity among the organizations interested in adopting this solution. First, they must adopt best practices for distributed identity as an additional application of their current security systems, including the ZTA model as a beneficial integration with distributed identity. Organizations cannot afford to implement different proprietary solutions with compatibility interfaces; thus, there is a need to conform to W3C DID standards across the different interoperability solutions. Furthermore, the possibility of technical and operating decentralization of identity needs to be analyzed, as well as the organization of training to ensure the proper functioning of this system. Human and ethical risks can be avoided if solutions follow privacy anticipations while abiding by regulations like the General Data Protection Regulation (GDPR) or the California Consumer Data Privacy Act (CCPA). Through such measures, organizations get maximum value out of distributed identity – guaranteed safe access without compromising integrated efficiency and compliance.

The potential of distributed identity for the modern cybersecurity change is quite large. Its full implementation also needs further analysis, innovation, and partnership among actors—especially in addressing technical, implementation, and policy challenges. It is necessary to conduct more extensive research on integrating distributed identity with ZTA improvements and look at integration concerns and application scaling. Furthermore, cooperation between companies and organizations to create rules for decentralized identity will be valuable when the technology is introduced to the broad public, and everyone will need to trust the decentralized digital environment. It is important to note that the shift towards more secure and protective identity management has started. Organizations, regulatory bodies, and technology innovators will need to go further in implementing true distributed identity as one of the critical cybersecurity components in the future.

# References

1. Jimmy, F. Emerging threats: The latest cybersecurity risks and the role of artificial intelligence in enhancing cybersecurity defenses. *Valley Int. J. Digit. Libr.* 2021, *564*, 574.
2. Daah, C.; Qureshi, A.; Awan, I.; Konur, S. Enhancing zero trust models in the financial industry through blockchain integration: A proposed framework. *Electronics* 2024, *13*(5), 865.
3. Dib, O.; Rababah, B. Decentralized identity systems: Architecture, challenges, solutions and future directions. *Ann. Emerg. Technol. Comput.* 2020, *4*(5), 19–40.
4. Liu, Y.; He, D.; Obaidat, M.S.; Kumar, N.; Khan, M.K.; Choo, K.K.R. Blockchain-based identity management systems: A review. *J. Netw. Comput. Appl.* 2020, *166*, 102731.
5. Rodný, P. SAML SSO Design. *Inf. Technol. Appl.* 2020, *9*(2), 55–62.
6. Fang, J.; Feng, T.; Guo, X.; Wang, X. Privacy-enhanced distributed revocable identity management scheme based self-sovereign identity. *J. Cloud Comput.* 2024, *13*(1), 154.
7. Buck, C.; Olenberger, C.; Schweizer, A.; Völter, F.; Eymann, T. Never trust, always verify: A multivocal literature review on current knowledge and research gaps of zero-trust. *Comput. Secur.* 2021, *110*, 102436.
8. Saxena, N.; Hayes, E.; Bertino, E.; Ojo, P.; Choo, K.K.R.; Burnap, P. Impact and key challenges of insider threats on organizations and critical businesses. *Electronics* 2020, *9*(9), 1460.
9. Mahdavifar, S.; Ghorbani, A.A. DeNNeS: Deep embedded neural network expert system for detecting cyber attacks. *Neural Comput. Appl.* 2020, *32*(18), 14753–14780.
10. Bhattacharya, M.P.; Zavarsky, P.; Butakov, S. Enhancing the security and privacy of self-sovereign identities on Hyperledger Indy blockchain. In *Proceedings of the 2020 International Symposium on Networks, Computers and Communications (ISNCC)*, Online, 1–7, 2020.
11. Lepore, C.; Laborde, R.; Eynard, J.; Kandi, M.A.; Macilotti, G.; Ferreira, A.; Sibilla, M. Assessing e-identity solutions according to self-sovereign identity: Application to eIDAS. *Asian Perspect.* 2023.
12. Van der Straaten, J. Identification for development it is not: 'Inclusive and trusted digital ID can unlock opportunities for the world's most vulnerable.' *SSRN Electron. J.* 2020.
13. Ghaffari, F.; Gilani, K.; Bertin, E.; Crespi, N. Identity and access management using distributed ledger technology: A survey. *Int. J. Netw. Manag.* 2022, *32*(2), e2180.
14. Muhammad, T.; Munir, M.T.; Munir, M.Z.; Zafar, M.W. Integrative cybersecurity: Merging zero trust, layered defense, and global standards for a resilient digital future. *Int. J. Comput. Sci. Technol.* 2022, *6*(4), 99–135.
15. Glöckler, J.; Sedlmeir, J.; Frank, M.; Fridgen, G. A systematic review of identity and access management requirements in enterprises and potential contributions of self-sovereign identity. *Bus. Inf. Syst. Eng.* 2024, *66*(4), 421–440.
16. Ugbebor, F.; Aina, O.; Abass, M.; Kushanu, D. Employee cybersecurity awareness training programs customized for SME contexts to reduce human-error-related security incidents. *J. Knowl. Learn. Sci. Technol.* 2024, *3*(3), 382–409.
17. Janssen, M.; Weerakkody, V.; Ismagilova, E.; Sivarajah, U.; Irani, Z. A framework for analyzing blockchain technology adoption: Integrating institutional, market, and technical factors. *Int. J. Inf. Manag.* 2020, *50*, 302–309.
18. Raskar, R.; Schunemann, I.; Barbar, R.; Vilcans, K.; Gray, J.; Vepakomma, P.; Sharma, V. Apps gone rogue: Maintaining personal privacy in an epidemic. *arXiv* 2020, 2003.08567.
19. Esposito, C.; Ficco, M.; Gupta, B.B. Blockchain-based authentication and authorization for smart city applications. *Inf. Process. Manag.* 2021, *58*(2), 102468.

20. Stafford, V. Zero trust architecture. *NIST Spec. Publ.* 2020, *800*, 207.
21. Soltani, R.; Nguyen, U.T.; An, A. A survey of self-sovereign identity ecosystem. *Secur. Commun. Netw.* 2021, *2021*(1), 8873429.
22. Ahmed, M.R.; Islam, A.M.; Shatabda, S.; Islam, S. Blockchain-based identity management system and self-sovereign identity ecosystem: A comprehensive survey. *IEEE Access* 2022, *10*, 113436–113481.
23. Martínez-Galán, E.; Leandro, F.J.B. A qualitative cost-benefit analysis of the maritime silk road in Europe: Who benefits from the initiative and who does not. *Asian Perspect.* 2024, *48*(1), 13–39.
24. Mazzocca, C.; Acar, A.; Uluagac, S.; Montanari, R.; Bellavista, P.; Conti, M. A survey on decentralized identifiers and verifiable credentials. *arXiv* 2024, 2402.02455.
25. Liu, Q.; Geertshuis, S.; Grainger, R. Understanding academics' adoption of learning technologies: A systematic review. *Comput. Educ.* 2020, *151*, 103857.
26. Khalid, M.I.; Ahmed, M.; Kim, J. Enhancing data protection in dynamic consent management systems: Formalizing privacy and security definitions with differential privacy, decentralization, and zero-knowledge proofs. *Sensors* 2023, *23*(17), 7604.
27. Duan, S.; Wang, D.; Ren, J.; Lyu, F.; Zhang, Y.; Wu, H.; Shen, X. Distributed artificial intelligence empowered by end-edge-cloud computing: A survey. *IEEE Commun. Surv. Tutor.* 2022, *25*(1), 591–624.
28. Halpin, H. Vision: A critique of immunity passports and W3C decentralized identifiers. *Secur. Stand. Res.* 2020, 148–168.
29. Xing, Y.; Lu, H.; Zhao, L.; Cao, S. Privacy and security issues in mobile medical information systems MMIS. *Mob. Netw. Appl.* 2024, 1–12.